\documentclass[runningheads]{llncs}

\usepackage{url}

\usepackage{array}
\usepackage{graphicx}
\usepackage{amsfonts} 
\begin{document}
\mainmatter              
\title{Out-of-Domain Semantics to the Rescue! Zero-Shot Hybrid Retrieval Models}

%
\titlerunning{Out-of-Domain Semantics to the Rescue}  
%


\author{Tao Chen \and  Mingyang Zhang \and Jing Lu \and Michael Bendersky \and Marc Najork}

\authorrunning{T. Chen et al.} 

\institute{Google Research, Mountain View, CA 94043, USA,\\
\email{\{taochen,mingyang,ljwinnie,bemike,najork\}@google.com}}

\maketitle              

\begin{abstract}

The pre-trained language model ({\it eg,} BERT) based deep retrieval models achieved superior performance over lexical retrieval models ({\it eg,} BM25) in many passage retrieval tasks. However, limited work has been done to generalize a deep retrieval model to other tasks and domains. In this work, we carefully select five datasets, including two in-domain datasets and three out-of-domain datasets 
with different levels of domain shift, and study the generalization of a deep model in a zero-shot setting. Our findings show that the performance of a deep retrieval model is significantly deteriorated when the target domain is very different from 
the source domain that the model was trained on. 
On the contrary, lexical models are more robust across domains. We thus propose a simple yet effective framework to integrate lexical and deep retrieval models. Our experiments demonstrate that these two models are complementary, even when the deep model is weaker in the out-of-domain setting.  The hybrid model obtains an average of 20.4\% relative gain over the deep retrieval model, and an average of 9.54\% over the lexical model in three out-of-domain datasets.



\keywords{deep retrieval, lexical retrieval, zero-shot learning, hybrid model}

\end{abstract}

\section{Introduciton}
\label{sec:introduction}


Traditionally, search engines have used lexical retrieval models ({\it eg,} BM25) to perform query-document matching. Such models are efficient and simple, but are vulnerable to vocabulary mismatch when queries use different terms to describe the same concept~\cite{Berger2000}. 
Recently, deep pre-trained language models ({\it eg,} BERT) have shown strong ability in modeling text semantics and have been widely adopted in retrieval tasks. Unlike lexical retrievers, deep/dense retrievers\footnote{While we recognize that in some cases the deep retrievers are not necessarily dense, and vice versa, we loosely use these two terms interchangeably throughout the paper.} capture the semantic relevance between queries and documents in a lower dimensional space, bridging the vocabulary mismatch gaps. Deep retrievers have been successful in many retrieval benchmarks. For instance, the most recent five winners in MS-MARCO passage~\cite{Bajaj2018} ranking leaderboard adopt deep retrievers as their first-stage retrieval model.

However, training a deep retrieval model is computationally expensive and a sizable labeled dataset to guide model training is not always available. A natural question then arises, can we train a deep retrieval model in one domain, and then directly apply it to new datasets/domains in a zero-shot setting with no in-domain training data? 
To answer this question, we carefully select five datasets, including two in-domain, 
and three out-of-domain datasets 
with different levels of domain shift. Through comprehensive experiments, we find that a deep retriever model performs well on related domains, but deteriorates when the target domain is distinct from the model source domain. On the contrary, lexical models are rather robust across datasets and domains. Our further analysis shows that lexical and deep models can be complementary to each other, retrieving different sets of relevant documents.

Inspired by this, we propose a zero-shot hybrid retrieval model to combine lexical and deep retrieval models. For simplicity and flexibility, we train a deep retrieval model and a lexical model separately and integrate the two (or more) models via Reciprocal Rank Fusion. This non-parametric fusion framework can be easily applied to any new datasets or domains, without any fine-tuning. Our experiments demonstrate the effectiveness of the hybrid model in both in-domain and out-of-domain datasets. 
In particular, though the zero-shot deep model is weaker in out-of-domain datasets, the hybrid model brings an average of 20.4\% of relative recall gain over the deep retrieval model, and an average of 9.54\% gain over lexical model (BM25) in three out-of-domain datasets. It also outperforms a variety of stronger baselines including query and document expansion.

To summarize, in this paper we explore the following research questions:
\begin{itemize}
    \item \textbf{RQ 1}: Can deep retrieval generalize to a new domain in a zero-shot setting?
    \item \textbf{RQ 2} Is deep retrieval complementary to lexical matching and query and document expansion? 
    \item \textbf{RQ 3}: Can lexical matching, expansion, and deep retrieval models be combined in a \emph{non-parametric hybrid retrieval} model?
\end{itemize}

To the best of our knowledge, this paper is the first to propose a hybrid retrieval model that incorporates lexical matching, expansion and deep retrieval in a zero-shot setup.  We demonstrate that the proposed hybrid model is simple yet effective in a variety of datasets and domains.


\section{Related Work}
\label{sec:related_work}

Information retrieval systems usually contain of two main stages: (a) \emph{candidate retrieval} (b) \emph{candidate re-ranking}. The retrieval stage is aimed at optimizing the recall of relevant documents, while the re-ranking stage optimizes early precision metrics such as NDCG@$k$ or MRR. Prior research ({\it eg,}~\cite{Bendersky2020}) found that the two stages are complementary -- gains in retrieval recall often lead to better early precision. Therefore, in this paper, we focus on retrieval recall optimization, with the assumption that the findings can benefit the re-ranking stage as well.

\textbf{Lexical retriever}
Traditionally, the first-stage retrieval has been a lexical based model such as  BM25~\cite{Robertson1994}, 
to capture the exact lexical match between queries and documents. Such simple and effective lexical models were the state-of-the-art for decades, and are still widely used in both academia and industry. One key issue with lexical models is the vulnerability to vocabulary mismatch, where queries and documents mention the same concept with different terms. One popular line to alleviate this is to expand terms in queries from pseudo relevance feedback ({\textit eg,} \cite{Jaleel2004} and \cite{Gianni2002}) or expand terms in documents from related documents ({\textit eg,} \cite{Tao2006}).
As a result, queries and documents have a higher chance to match each other at the surface form.

\textbf{Deep LM augmented lexical retriever}
More recently, pre-trained deep language models (LM) such as BERT~\cite{Devlin2019} have been shown to be powerful in many natural language understanding tasks. The very first application of such models in IR is to augment lexical retrieval models. Dai et al.~\cite{Dai2019,Dai2020} proposed to learn context-aware term weights by BERT to replace the term frequencies used by lexical models.  To remedy the vocabulary gap between queries and documents, Nogueira and Lin~\cite{Nogueira2019,Nogueira2019b} employed seq2seq model transformer~\cite{Vaswani2017} and later T5~\cite{Raffel2020} to generate document expansions, which brings significant gains for BM25. In the same vein, Mao et al.~\cite{Mao2021} adopted seq2seq model BART~\cite{Lewis2020} to generate query expansions, which outperforms RM3~\cite{Jaleel2004}, a highly performant lexical query expansion method.

\textbf{Deep retriever}
In a separate line of research, deep neural retrieval models adopt LMs to build a new paradigm for first-stage retrieval:  instead of performing exact lexical match, they aim at capturing the relevance of queries and documents in a lower dimensional semantic space. This paradigm can largely bridge the vocabulary gap between queries and documents. Since cross-attention models are cost-prohibitive for first-stage retrieval, most works adopt a dual-encoder  architecture to learn two single vector representations for the query and the document separately, and then measure their relevance by a simple scoring function  (\textit{eg,} dot product or cosine similarity). In this way, finding most relevant documents can be formulated as a nearest neighbor search problem and can be accelerated with quantization techniques~\cite{Johnson2021,Guo2020}. 

For model training, it is often the case that positive (query, document) pairs are available, while negative pairs need to be sampled from the dataset. Negative sampling strategy plays a crucial role for model performance. 
Earlier works adopt simple in-batch negative sampling~\cite{Zhan2020,Kuzi2020}, or mine negative pairs from top BM25 results~\cite{Karpukhin2020}. Recent works propose more sophisticated sampling strategies to identify high-quality hard negatives, such as cross-batch negatives~\cite{Qu2021}, demonised hard negatives~\cite{Qu2021} and semantic similarity 
based negatives~\cite{Lu2021}.

Deep retriever model has shown superior performance over lexical models in several passage retrieval tasks ({\textit eg,} MS-MARCO passage ranking~\cite{Bajaj2018}). 
However, training a deep model is expensive computationally but also in terms of labeled data creation. A simple remedy is to directly apply a trained deep retriever model to new domains in a zero-shot setting. However, 
little work has been conducted to uncover the generalization ability of deep retrievers. 
One exception is by Thakur et al. \cite{Thakur2021} who introduce BEIR, an IR benchmark of 18 datasets with diverse domains and tasks, and evaluate several trained deep models in a zero-shot setup. They found that deep models exhibit a poor generalization ability, and are significantly worse than BM25 on datasets that have a large domain shift compared from what they have been trained on. In our work, we conduct similar studies, and observe the same performance deterioration for a deep model in zero-shot setting. We additionally propose a hybrid model to utilize a lexical model to alleviate the domain shift.

\textbf{Hybrid retriever}
Deep retrievers are good at modeling semantic similarity, while could be weaker at capturing exact match or could have capacity issues when modeling long documents~\cite{Luan2021,Wang2021b}. A few recent works attempt to build a hybrid model to take the strength of both deep and lexical retrievers. Most works train a deep model separately and then interpolate its score with a lexical model score~\cite{Wang2021b,Karpukhin2020,Luan2021,Lin2021,Ma2021b,Lin2021brief}, or use RM3 built on the top lexical results to select deep retriever results as the final list~\cite{Kuzi2020}, or simply combine the results of the two models in an alternative way~\cite{Zhan2020}. Gao et al.~\cite{Gao2021} is the only work that explicitly trains a deep model to encode semantics that lexical model fails to capture. In model inference, they interpolate the scores of deep and lexical models and generate the top retrieval results. While insightful, these prior works limit the model evaluation to a single task and a single domain. It is unclear how such hybrid model performs in a cross-domain setting, without any fine-tuning. Our work aims to fill this research gap, and demonstrates that a zero-shot hybrid retrieval model can be more effective than either of the two models alone.









\section{Method}
\label{sec:method}

In this section, we describe our zero-shot hybrid retrieval model. For simplicity and flexibility, we train deep and lexical retrieval models separately, and propose a simple yet effective non-parametric framework to integrate the two.

\subsection{Hybrid retrieval model}
Both traditional lexical retrieval models~\cite{Robertson1994,Ponte+Croft:1998,Gianni2002}, as well as deep neural retrieval models~\cite{Gao2021,Kuzi2020,Wang2020} represent queries and documents using vectors $\mathbf{q}, \mathbf{d} \in \mathbb{R}^N$, and score candidates based on the dot product $<\mathbf{q}, \mathbf{d}>$. Thus, the difference between deep and lexical models stems from how these vectors are constructed. 

Lexical models represent queries and documents using sparse weight vectors $\mathbf{q}^{sparse}, \mathbf{d}^{sparse} \in \mathbb{R}^V$, respectively (where $V$ denotes the vocabulary size). The vectors are sparse such that all the entries for vocabulary terms that do not appear in query and document are zeroed out. To combat issues of term mismatch, lexical models often include additional terms in queries and document through some form of expansion (\textit{eg,} based on pseudo-relevance feedback~\cite{Lavrenko+Croft:}). However, the resulting vectors are still highly sparse, due to the high dimensionality of vocabulary size $V$.

In contrast, deep neural retrieval models represent queries and documents using dense embedding vectors $\mathbf{q}^{dense}, \mathbf{d}^{dense} \in \mathbb{R}^E$, where $E << V$. While theoretically dense embeddings overcome the term mismatch problem, they do have several shortcomings. First, they require large amounts of data and resources for training~\cite{Qu2021}, and thus may not be directly trained over collections with fewer queries and relevance judgments. Second, they do not capture \emph{exact} query-document matches as well as the sparse lexical scores. Therefore, a lexical and deep model combination is likely to yield the optimal relevance scores.

Most prior works~\cite{Wang2021b,Karpukhin2020,Luan2021,Lu2021} model this combination as a linear interpolation of the scores of deep and lexical retrieval models. This fusion method is sensitive to the score scales and the weights assigned to the different models~\cite{Wang2021b}, which needs careful score normalization and weight tuning, especially when multiple models are combined. We expect that the raw scores of the models can vary from one domain/dataset to another, and likewise the interpolation weights.

Since our goal is to build a hybrid model which can be easily applied to a new domain in a zero shot setting (with no in-domain training data), we would like to eliminate such domain-specific normalization and tuning. Therefore, we adopt Reciprocal Rank Fusion (RRF)~\cite{Cormack2009} to generate the final ranking results by considering the \emph{ranking positions} of each candidate generated by different models, instead of fusing their scores. RRF demonstrates robust and effective ensembles in prior works~\cite{Bendersky2020,Cormack2009} and our experiments. Assuming a set of lexical and deep retrieval models $M$, we define $\pi^m(q,d)$ as the rank for document $d$, induced by its score for query $q$ assigned by model $m \in M$. The RRF score is then defined as:
\begin{equation}
    RRF(q, d, M) = \sum_{m \in M}\frac{1}{k + \pi^m(q,d)}
    \label{eq:rrf_score}
\end{equation}
where $k = 60$, following the definition in the original paper~\cite{Cormack2009}.

In the remainder of this paper we demonstrate that this simple non-parametric approach generalizes well across domains, and can make an effective use of out-of-domain semantics of retrieval models trained on a different collection. In the remainder of this section, we describe the lexical and deep retrieval models used to instantiate Equation~\ref{eq:rrf_score}.

\subsection{Lexical retrieval model}
We adopt \textbf{BM25} as the base lexical retrieval model, as it is widely used and shown to be robust~\cite{Thakur2021}. To alleviate the vocabulary mismatch issue, we additionally apply popular query expansion and document expansion techniques to expand the query and the document, forming enhanced lexical models. 

\textbf{BM25+Query expansion} 
Most conventional query expansion approaches follow the pseudo-relevance feedback (PRF) paradigm. It assumes the top K ranked documents for the original query to be relevant, and generates query expansions from these documents. In our work, we experiment with \textbf{RM3}~\cite{Jaleel2004} (a relevance-based language model) and \textbf{Bo1}~\cite{Gianni2002} (a variant of Divergence From Randomness term weighting model), to obtain query expansions from PRF. 

\textbf{BM25+Document expansion}
Recently, generative models like T5 were shown to generate high-quality document expansions, and bring large gains to the BM25 model on retrieval tasks~\cite{Nogueira2019,Nogueira2019b,Pradeep2021}. Following the \textbf{docT5query} approach~\cite{Nogueira2019b,Pradeep2021}, we fine-tune T5-base with identical setting as the prior works on (query, relevant passage) pairs from the MS-MARCO passage ranking training set, where the query is considered as pseudo document expansion. We adopt the top-$k$ sampling decoder~\cite{Fan2018} to generate $N$ (a tunable parameter) queries per passage. For each document, we append the expansions to each passage and aggregate them as the document expansion. 


\subsection{Deep retrieval model}
We adopt \textbf{NPR}~\cite{Lu2021}, a neural passage retrieval model with improved negative contrast as the deep retrieval model in our framework. Note that our framework is flexible, and NPR can be replaced with any other deep model. Aligned with many popular deep retrievers \cite{Karpukhin2020,Qu2021,ance},
NPR adopts a dual encoder architecture, learning dense embedding vectors representations, computing the relevance using the dot product $<\mathbf{q}^{dense}, \mathbf{d}^{dense}>$. The training of this model is enhanced with several negative sampling strategies, aiming at obtaining hard and high-quality negative (query, passage) pairs.  This model is trained on MS-MARCO passage dataset (detailed in Section \ref{subsec:dataset}), and achieves a very competitive performance. 
To adapt NPR to document retrieval setting, we split documents into passages by applying sliding overlapping sentence windows. Following work by 
\cite{dai2019_deeper}, we use the max passage retrieval score as the document level score.
\section{Experimental Setup}
\label{sec:experiment}

\subsection{Datasets}
\label{subsec:dataset}

As we are interested in exploring the performance of the deep retrieval model in a variety of out-of-domain settings, we choose to specifically focus on five datasets in our evaluation (summarized in Table~\ref{tab:datasets}).
\begin{enumerate}
    \item \textbf{MS-MARCO passage}~\cite{Bajaj2018} dev set is the dataset we use for the \emph{in-domain} model evaluation, as the NPR deep retrieval model, and the docT5query model are trained using the training portion of this dataset (see Section~\ref{subsec:processing} for more details). The queries in this dataset are all questions.
    \item \textbf{MS-MARCO doc}~\cite{Bajaj2018} 
    is derived from MS-MARCO passage, but instead the retrieval is done using documents. We use the queries in dev set for evaluation (a subset of MS-MARCO passage dev set). This evaluates the generalization of the model to document retrieval.
    \item \textbf{ORCAS}~\cite{Craswell2020} is a click dataset based on an intersection of Bing search engine logs and the documents in MS-MARCO dataset. Compared to MS-MARCO, queries in ORCAS exhibit wider topics (not limited to questions) and shorter length
    (76\% of queries have no more than 3 tokens after removing stopwords). Since it has a very large number of queries (10M), we evaluate our model using a stratified sample of 10k queries, based on query length.
    \item \textbf{Robust04}~\cite{Voorhees2005} is a dataset comprising 528K news stories and 250 queries. 
    Each query consists of three fields, including title (keywords), description (a sentence-length statement of the information needs) and narrative (a paragraph-length text explaining what makes a document relevant). 
    It evaluates how well the retrieval model generalizes to the news domain.
    \item \textbf{TREC-COVID}~\cite{Roberts2020} is based on the CORD19~\cite{Wang2020} collection -- PubMed articles and preprints about the COVID-19 pandemic. Each query contains a few keywords, along with a more specific natural language version of question, and a narrative which adds additional clarifications of user intent. As shown by Thakur et al.~\cite{Thakur2021}, it is quite distinct from the MS-MARCO dataset, and provides a good test case for whether an out-of-domain retrieval system can be useful in a bio-medical domain. 
    
\end{enumerate}

\begin{table}[t]
\caption{The five datasets used for model evaluation. ``Avg. D/Q'' denotes the average number of relevant docs per query. \label{tab:datasets}}
\begin{tabular}{l|l|l|r|r|r}
\hline
Dataset           & Domain      & Task              & \multicolumn{1}{l|}{\#Query} & \multicolumn{1}{l|}{\#Corpus} & \multicolumn{1}{l}{Avg. D/Q} \\ \hline
MS-MARCO passage~\cite{Bajaj2018}  & Misc.       & Passage retrieval & 6980                          & 8.8M                          & 1.1                           \\ \hline
MS-MARCO doc~\cite{Bajaj2018} & Misc        & Doc retrieval     & 5193                          & 3.2M                          & 1.1                           \\ \hline
ORCAS~\cite{Craswell2020}             & Misc.       & Doc retrieval     & 9670                          & 1.4M                          & 1.8                           \\ \hline
Robust04~\cite{Voorhees2005}          & News        & Doc retrieval     & 250                           & 528K                          & 69.9                          \\ \hline
TREC-COVID~\cite{Roberts2020}        & Bio-medical & Doc retrieval     & 50                            & 191K                          & 493.5                         \\ \hline
\end{tabular}
\end{table}


\subsection{Data processing and benchmarking}
\label{subsec:processing}

In following, we detail our experimental setup to ensure the reproducibility of all the reported results.

\textbf{Deep retrieval model} As described in Section~\ref{sec:method}, we train NPR on the training set of MS-MARCO passage dataset, and apply this model to the other four datasets without any fine-tuning. The documents in the other four datasets are long and may exceed the 512 token length limitation.  Following prior work~\cite{Pradeep2021}, we use a sliding window of ten sentences with a stride of five to split each document into passages. We run NPR on each passage, perform the nearest neighbor search via SCaNN~\cite{Guo2020} at passage-level and consider the best passage score as its document score. The query used for each dataset is the same as BM25 based lexical model (detailed in Table~\ref{tab:lexical-model-setup}).

\begin{table}[th]
\caption{The best setup for lexical retrieval models. ``des./narr./ques.'' denotes description/narrative/question field and ``\#fk docs/terms'' denotes the number of feedback documents/terms. }
\label{tab:lexical-model-setup}
\begin{tabular}{l|l|l|l|l|l}
    Model  ($\rightarrow$)              & \multicolumn{2}{c|}{BM25}     & \multicolumn{2}{c|}{Bo1} & \multicolumn{1}{c}{docT5query} \\ \hline
    Dataset ($\downarrow$)              & index     & query & \#fk doc   & \#fk terms  & \#expansions                         \\ \hline
MS-MARCO passage  & full text & query             & 5          & 10          & 40                              \\
MS-MARCO doc & full text & query             & 5          & 5           & 20/passage                      \\
ORCAS             & full text & query             & 10         & 10          & 20/passage                      \\
Robust04          & full text & query+des.+narr.  & 5          & 10          & 10/passage                      \\
TREC-COVID        & abstract  & query+ques.+narr. & 20         & 40          & 10                             
\end{tabular}
\end{table}

\textbf{Lexical retrieval models} For implementing our lexical models, we use the Terrier search engine~\cite{Macdonald2012}, and apply the default options for stemming and stop word removal provided by Terrier. We employ three fully lexical benchmarks. We carefully tune the parameters, and detail the settings in Table~\ref{tab:lexical-model-setup}.
\begin{itemize}
    \item BM25 is a commonly used bag-of-words retrieval method. We use the default parameters provided by Terrier, and verify that our results (in terms of MAP) are comparable to other previously reported BM25 benchmarks~\cite{Yang+al:2018}. We experiment with a few indexing options: 1) full text, 2) passage and 3) abstract for TREC-COVID only.
    \item Bo1 is a query expansion package implemented in Terrier. 
    For each dataset, we carefully tune the number of feedback documents ([5, 10, ..., 50]) and the number of feedback terms (\textit{ie}, expansions; [5, 10, ..., 60]). We also experiment with RM3 query expansion package by Terrier and carefully tune the two parameters. However, it yields lower performance than Bo1 in all the five datasets. We thus only report the results of Bo1 in the Section~\ref{sec:results}.
    
    \item docT5query is a T5 based document expansion model. As described in Section~\ref{sec:method}, we fine-tune T5 model on the MS-MARCO passage training set by strictly following the setup of prior works~\cite{Nogueira2019b,Pradeep2021}. We feed each passage length text, namely, passage in the MS-MARCO passage collection, the abstract in TREC-COVID, or split passages of other three datasets, to T5 model and generate N (a tunable parameter; [10, 20, 40]) numbers of expansions. We append the expansions for all the passages to a document. 
\end{itemize}


\section{Evaluation}
\label{sec:results}

As our work focuses on the first stage retrieval, in this section we adopt Recall@1K as the primary evaluation metric and additionally report MAP score. In our evaluation, we aim to address the research questions posed in Section~\ref{sec:introduction}.

\subsection{Generalization of the deep retrieval model}
\label{subsec:deep-model-result}

We first focus on the results on two in-domain datasets (Table~\ref{tab:results-in-domain}).
As expected, the deep retrieval model NPR performs very well on MS-MARCO passage on which it is trained. It substantially beats BM25 by an absolute 10.77 (relative 12.35\%) and 16.15 (relative 83.59\%) in terms of Recall@1K and MAP, respectively. In MS-MARCO doc (the in-domain document retrieval task), NPR also performs well, and betters BM25 by 4.55 (5.0\%) and 3.86 (14.57\%) at Recall@1K and MAP, respectively. This indicates that a well-trained deep passage retrieval model generalizes well to an in-domain document retrieval task.


\begin{table}[th]
\centering
\caption{Experimental results on two in-domain datasets. The improvements (R@1K) of all hybrid models (5-8) over baselines (1-4) are statistically significant via a paired two-tailed t-test ($p<0.05$). 
}
\label{tab:results-in-domain}
\begin{tabular}{l|>{\centering\arraybackslash}p{0.15\textwidth}|>{\centering\arraybackslash}p{0.15\textwidth}|>{\centering\arraybackslash}p{0.15\textwidth}|>{\centering\arraybackslash}p{0.15\textwidth}}

Dataset ($\rightarrow$)        & \multicolumn{2}{c|}{MS-MARCO passage}                & \multicolumn{2}{c}{MS-MARCO doc}                \\ \hline
Model ($\downarrow$)             & \multicolumn{1}{c|}{R@1K} & \multicolumn{1}{c|}{MAP} & \multicolumn{1}{c|}{R@1K} & \multicolumn{1}{c}{MAP}  \\ 
\hline
1. BM25            & 87.18                     & 19.32                    & 90.91                     & 26.50                    \\
2. BM25+Bo1        & 88.27                     & 17.95                    & 91.64                     & 22.69                    \\
3. BM25+docT5query & 94.07                     & 26.09                    & 93.18                     & 30.28                    \\ 
\hline
4. NPR             & 97.95                     & \textbf{35.47}                    & 95.46                     & 30.36                    \\ 
\hline
5. RRF(1, 4)       & 98.31                     & 29.46                    & 96.80                     & 32.09                    \\
6. RRF(2, 4)        & 98.36                     & 28.62                    & 96.90                     & 31.48                    \\
7. RRF(3, 4)        & \textbf{98.65}            & 32.89                    & 96.86                    & \textbf{33.50}                    \\
8. RRF(2, 3, 4)   & 98.48	                 &29.58	                       & \textbf{96.96}	                  &32.48
\end{tabular}
\end{table}

\begin{table}[th]
\centering
\caption{Experimental results on three out-of-domain datasets. The improvements (R@1K) of all hybrid models (5-8) over baselines (1-4) are statistically significant via a paired two-tailed t-test ($p<0.05$), except 5/7 vs. 2 in Robust04 and TREC-COVID. 
}
\label{tab:results-out-of-domain}
\begin{tabular}{l|>{\centering\arraybackslash}p{0.1\textwidth}|>{\centering\arraybackslash}p{0.1\textwidth}|>{\centering\arraybackslash}p{0.1\textwidth}|>{\centering\arraybackslash}p{0.1\textwidth}|>{\centering\arraybackslash}p{0.1\textwidth}|>{\centering\arraybackslash}p{0.1\textwidth}}

    Dataset ($\rightarrow$)                & \multicolumn{2}{c|}{ORCAS}                          & \multicolumn{2}{c|}{Robust04}                       & \multicolumn{2}{c}{TREC-COVID}   \\ \hline
        Model ($\downarrow$)            & \multicolumn{1}{c|}{R@1K} & \multicolumn{1}{c|}{MAP} & \multicolumn{1}{c|}{R@1K} & \multicolumn{1}{c|}{MAP} & \multicolumn{1}{c|}{R@1K} & \multicolumn{1}{c}{MAP}   \\ \hline
1. BM25             & 77.52                    & 27.1                    & 72.84                    & 26.91                   & 49.29                    & 27.86 \\
2. BM25+Bo1       & 78.85                    & 23.53                   & 79.02                    & 30.83                   & 52.58                    & 30.98 \\
3. BM25+docT5query & 79.62                    & 30.28                   & 74.64                    & 28.01                   & 50.66                    & 28.77 \\ \hline
4. NPR              & 81.18                    & 28.29                   & 70.28                    & 28.39                   & 37.58                    & 17.14 \\ \hline
5. RRF(1, 4)        & 85.95                    & 30.33                   & 79.62                    & 33.19                   & 52.32                    & 30.38 \\
6. RRF(2, 4)        & 86.18                    & 28.36                   & \textbf{82.82}            & \textbf{34.60}                    & 54.63                    & 32.21 \\
7. RRF(3, 4)        & 86.44                    & \textbf{31.39}                   & 79.81                    & 33.34                   & 53.01                    & 30.64 \\
8. RRF(2, 3, 4)  & \textbf{86.49}	    &29.74	&82.65	&34.51	& \textbf{55.66}	& \textbf{34.22}
\end{tabular}
\end{table}

In Table~\ref{tab:results-out-of-domain}, we discuss the results of three out-of-domain document retrieval datasets. Compared to MS-MARCO doc, ORCAS dataset has the least domain shift (as the candidate documents stem from MS-MARCO doc albeit with different queries), followed by Robust04 (news domain). TREC-COVID contains COVID-19 specific topics and has the largest domain shift. We observe that NPR has a clear performance drop with the increased domain shift. NPR performs reasonably on ORCAS, and betters BM25 by relative 4.72\% and 4.39\% at Recall@1K and MAP, respectively. However, it still has an absolute drop of 14.28 and 2.07 in terms of Recall@1K and MAP, compared to its performance on MS-MARCO doc. 
In the news domain (Robust04 dataset), the performance of NPR is mixed: it outperforms BM25 by  5.5\% of relative MAP improvement, but underperforms by relative 3.51\% at Recall@1K.  In TREC-COVID dataset, BM25 significantly beats NPR by 11.71 (23.76\%) and 10.72 (38.48\%) in terms of Recall@1K and MAP. This demonstrates that the generalization ability of deep retrieval models is poor, especially when the target domain is dramatically different from its training domain.  

\subsection{Utility of query and document expansion }
Lexical retrieval models are prone to vocabulary mismatch between queries and documents. We examine whether query and document expansion models could bridge this gap. From Table~\ref{tab:results-in-domain} and Table~\ref{tab:results-out-of-domain}, we see that Bo1 query expansion model consistently brings recall gains, with $1\%$ relative gain on MS-MARCO passage/doc and ORCAS, 8.48\% on Robust04 and 6.67\% on TREC-COVID. 

Recall that docT5query document expansion model is trained on the training set of MS-MARCO passage dataset. In this dataset, it brings very large gains to BM25. In the other four datasets, docT5query shows a consistent, albeit smaller, improvement over BM25 (above 2.5\% recall gain), similar to the analysis by Thakur et al.~\cite{Thakur2021}. 


\subsection{Complementarity of lexical and deep retrieval models}

\begin{figure}[th]
    \centering
    \begin{tabular}{cc}
    \includegraphics[scale=0.55]{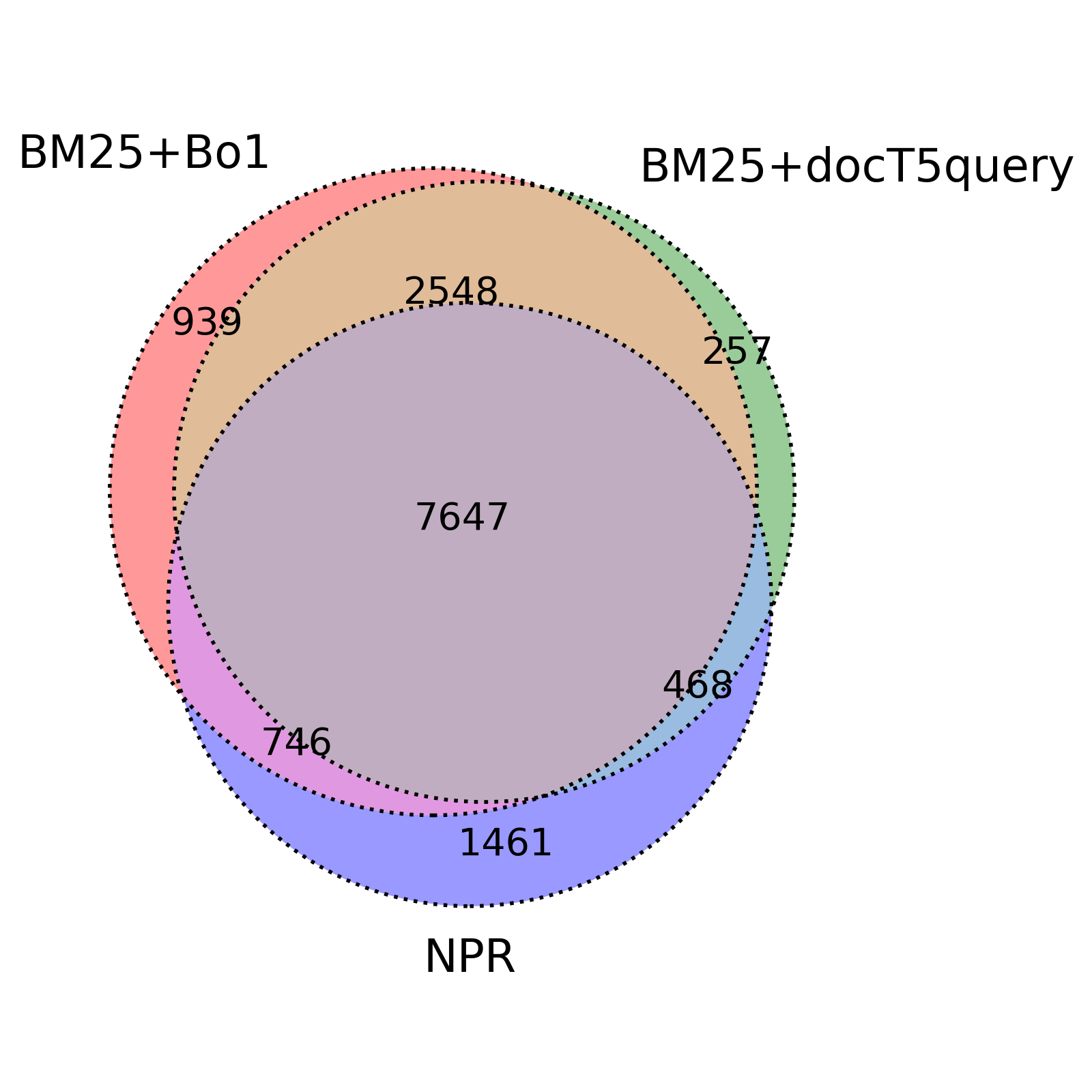}     &  
    \includegraphics[scale=0.55]{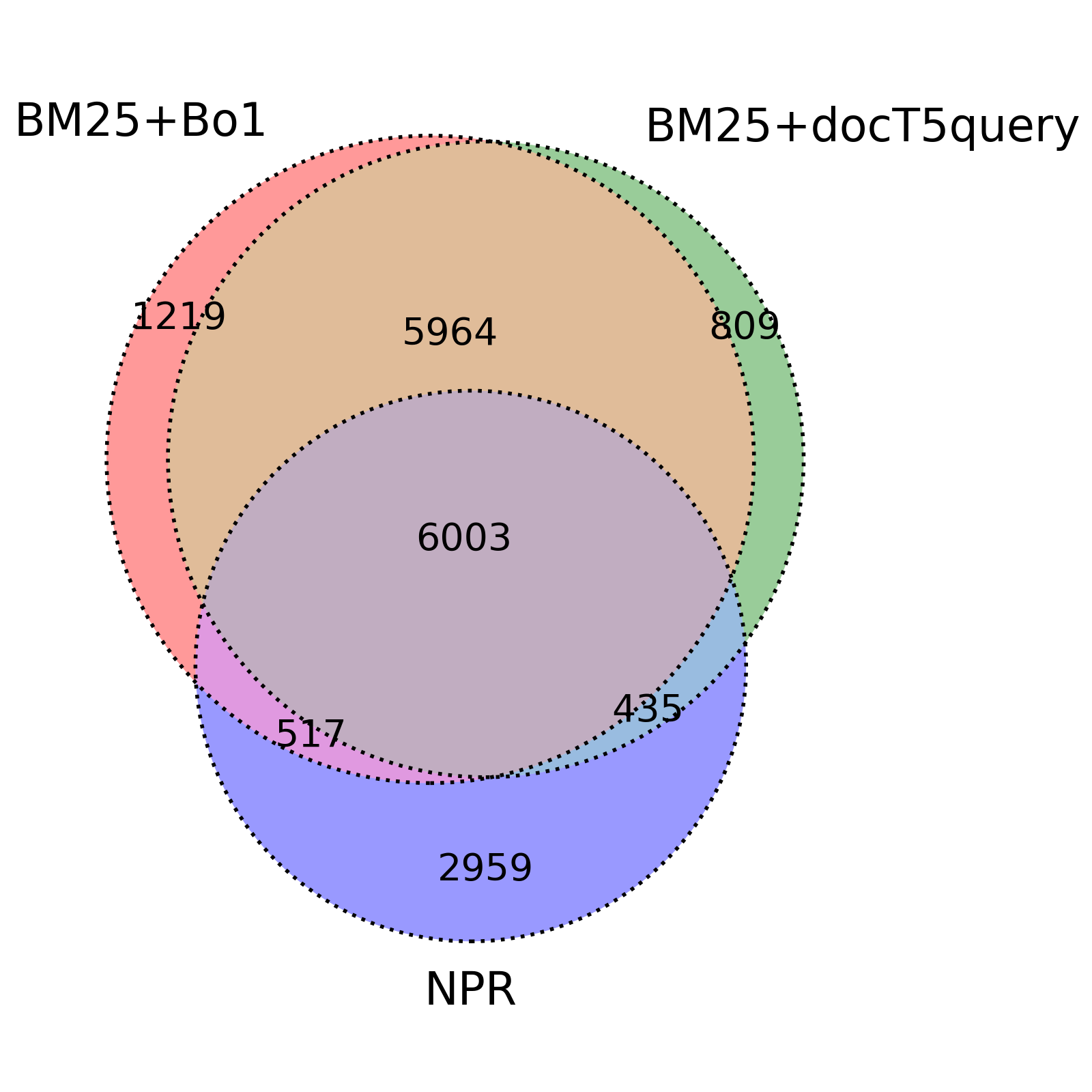} \\
    (a) Robust04 & (b) TREC-COVID
    \end{tabular}
    \caption{A Venn diagram of relevant results 
    by the Bo1, docT5query, and NPR. 
    }
    \label{fig:deep-lexical-venn}
\end{figure}

As with query/document expansion, deep retrieval model can narrow the vocabulary gap between queries and documents. One natural question is, are these models still complementary to each other?  To answer this, we plot the unique relevant documents retrieved by BM25+Bo1, BM25+docT5query and NPR and their overlaps in Figure~\ref{fig:deep-lexical-venn} for Robust04 and TREC-COVID (other three datasets only have around one relevance document per query, ref Table~\ref{tab:datasets}). We see that each method is complementary to each other. 
In general, NPR retrieves the largest number of unique relevant results, though it retrieves less relevant results than the other two methods.

\subsection{Effectiveness of the proposed hybrid model}
Our proposed hybrid framework provides a flexible mechanism for fusing multiple lexical or deep retrieval models. In Table~\ref{tab:results-in-domain} and Table~\ref{tab:results-out-of-domain} (row 5-8), we demonstrate the performance of our hybrid model which consistently outperforms either the lexical or deep retrieval model alone. In in-domain MS-MARCO passage dataset, the best performing hybrid model of BM25+docT5query and NPR (\#7) obtains a Recall@1K of 98.65, betters BM25 and NPR by relative 12.94\% and 0.52\%, respectively. This hybrid model outperforms coCondenser (Recall@1K=98.4)~\cite{Gao2021b}, the current MS-MARCO leaderboard winner (as of 2021/08/09) in the passage retrieval task.\footnote{Note that we focus solely on recall, since we do not apply a second re-ranking stage for optimizing early precision.} In the in-domain document retrieval task, the best performing hybrid model is the one with all the three methods (Bo1, docT5query and NPR).  

In three out-of-domain datasets, the advantage of hybrid model is more evident, given that NPR is weakened in datasets with a large domain shift (ie, TREC-COVID). It consistently improves over BM25 by almost 10\% relatively for the three datasets, and substantially outperforms NPR by 6.11\%, 14.16\% and 44.25\% in ORCAS, Robust04, and TREC-COVID, respectively. This demonstrates that our proposed zero-shot hybrid retrieval model is effective and robust across different tasks and domains. 

\begin{figure}[t]
    \centering
    \begin{tabular}{cc}
    \includegraphics[scale=0.35]{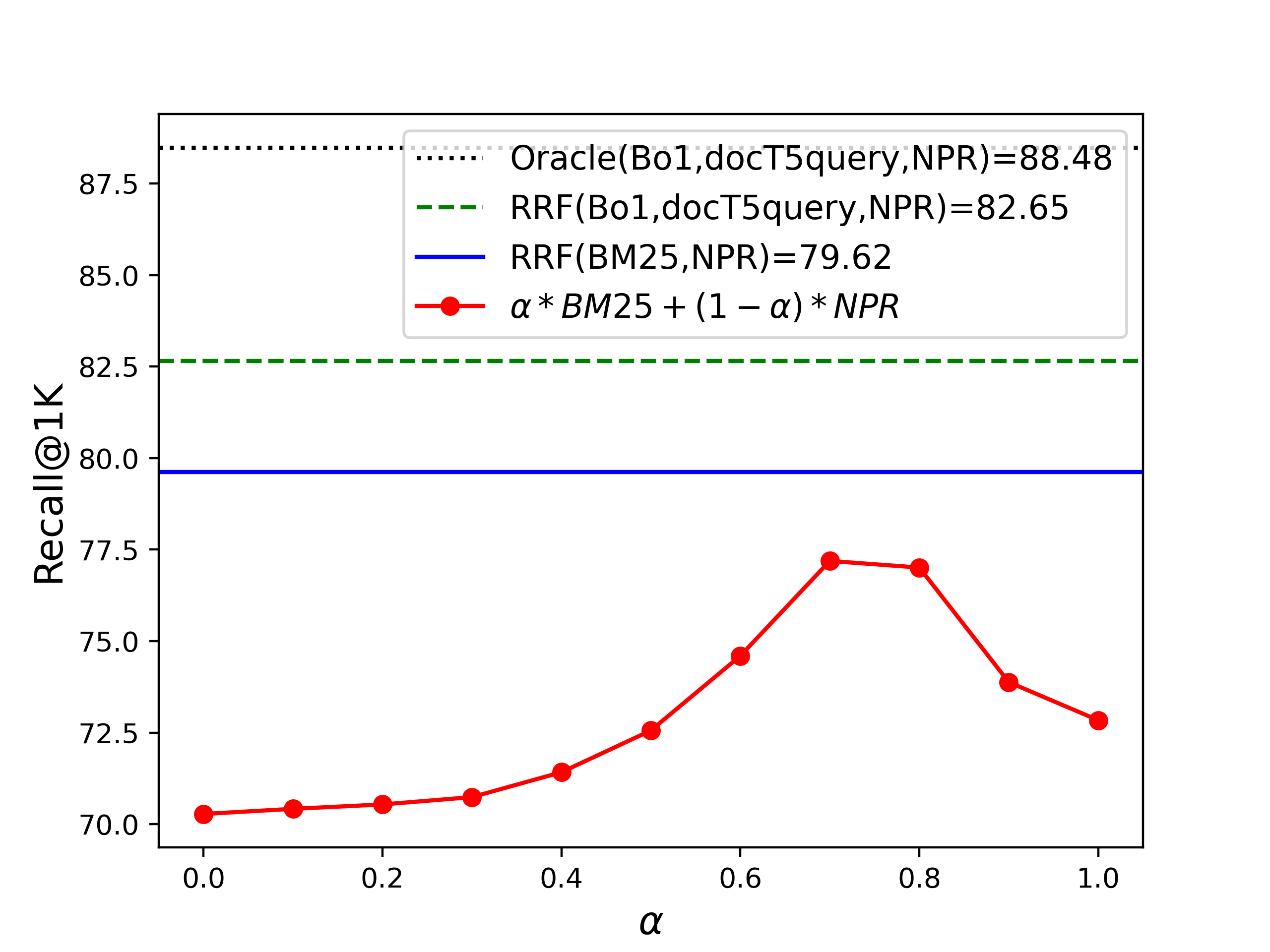}     &  
    \includegraphics[scale=0.35]{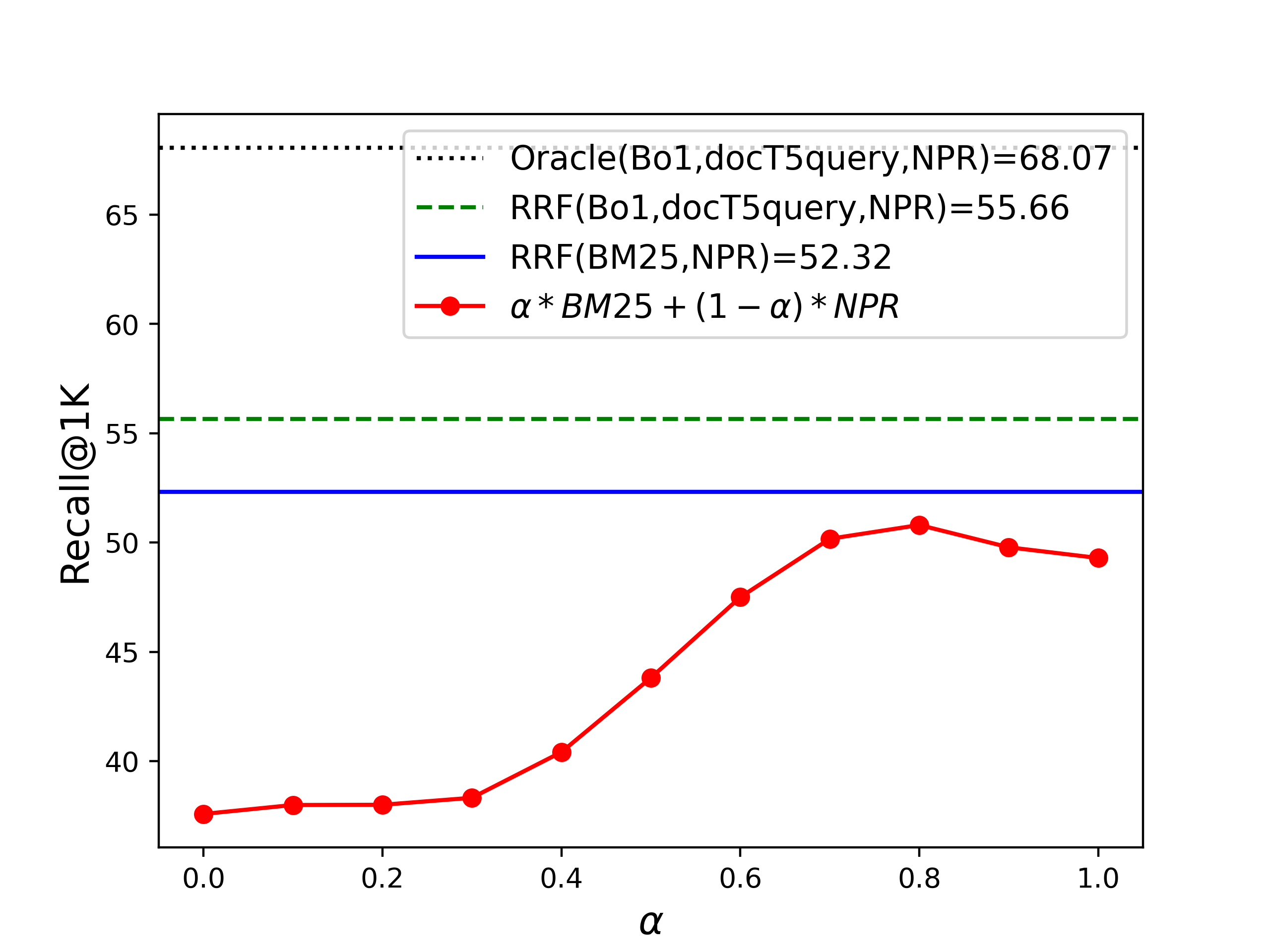} \\
    (a) Robust04 & (b) TREC-COVID
    \end{tabular}
    \caption{The comparisons of our hybrid model, oracle system and interpolation. 
    }
    \label{fig:system-comparison}
\end{figure}
\vspace{-4mm}

\section{Discussion}
\label{sec:discussion}

Our zero-shot hybrid model has demonstrated its effectiveness in the experiments. For comparison, we implement the linear interpolation method that most prior works adopted~\cite{Wang2021b,Karpukhin2020,Luan2021,Lu2021}, though such model is not zero-shot, and requires weight tuning. As weight tuning complexity increases with the number of models, we only interpolate BM25 and NPR as a case study: $s(d) = \alpha \times s_{BM25}(d) + (1-\alpha) \times s_{NPR}(d)$. We perform min-max score normalization and carefully tune the weight $\alpha \in [0.1, \ldots, 0.9]$ via grid search for out-of-domain datasets Robust04 and TREC-COVID.


Figure~\ref{fig:system-comparison} (bottom curve), shows that interpolation is weight-sensitive, and furthermore even the best setting underperforms our simple non-parametric hybrid model RRF(BM25, NPR) by a relative 3\% in both datasets. The differences are even larger, when compared with the full RRF model (dashed line).
We also explore the hybrid upper bound by fusing the retrieval results of BM25+Bo1, BM25+docT5query and NPR via an oracle, \textit{ie,} merging all relevant results from each method regardless of their ranking positions. Figure~\ref{fig:system-comparison} (dotted top line) illustrates the large potential headroom for designing an even better fusion model. 


Similarly to us, Wang et al.~\cite{Wang2021b} found that setting an oracle per-query weight yields better performance than optimizing a global weight. 
Inspired by this, we hypothesize that the performance of retrieval models relate to query length.
We bin the ORCAS queries into 10 groups, based on the number of non-stopword tokens, and show the breakdown results in Table~\ref{tab:orcas-length-result}. When the queries are very short, NPR largely beats BM25, even with query and document expansion.
However, its performance deteriorates for longer queries, with 7 or more tokens.

To gain more insights, we spot-check wins and losses. For single token queries, BM25 performs badly when the query is misspelled (\textit{eg}, ``ihpone6'') or a compound word (\textit{eg}, ``tvbythenumbers''). These words
are very likely to be out-of-vocabulary (OOV) in lexical retrieval models. On the contrary, deep retrieval model NPR adopts wordpiece tokenizer, which could still capture the semantics of the OOV from its sub-units. For long queries, NPR performs poorly for those employing complex logic and seeking very specific information, \textit{eg}, ``according to piaget, which of the following abilities do children gain during middle childhood?''. In this example query, BM25 successfully retrieves relevant documents containing the identical query sentence, while NPR fails. This may indicate that NPR is worse at capturing exact match, consistently with prior work~\cite{Wang2021b,Luan2021}.

\begin{table}[t]
\centering
\caption{Mean R@1K result by query length for ORCAS dataset (best result bolded).} 
\label{tab:orcas-length-result}
\small
\begin{tabular}{l|r|r|r|r|r|r|r|r|r|r}
Model \textbackslash Query Length                & 1 & 2 & 3 & 4 & 5 & 6 & 7 & 8 & 9 & 10 \\ \hline
1. BM25              & 35.0                 & 74.2                 & 80.2                & 82.0                 & 82.3                 & 82.9                & 81.1                 & 87.8                 & 85.6                 & 87.5                 \\
2. BM25+Bo1        & 39.5                 & 76.6                 & 81.3                 & 83.1                 & 83.3                 & 83.0                & 82.6                & 88.1                 & \textbf{86.3}                 & \textbf{87.9}                  \\
3. BM25+docT5query & 36.9                 & 77.8                & 83.0                 & 84.1               & 86.2                & 84.9                & \textbf{83.8}                 & \textbf{88.6}                 & 85.8                & 87.8                  \\ \hline
4. NPR               & \textbf{59.8}                 & \textbf{82.5}                 & \textbf{85.8}                 & \textbf{86.6}                 & \textbf{87.6}                 & \textbf{85.6}                 & 82.8               & 83.3                 & 80.1               & 75.9                 
\end{tabular}
\end{table}
\vspace{-4mm}




\section{Conclusion}
\label{sec:conclusion}

Compared to traditional lexical retrieval models, a deep retrieval model mitigates the vocabulary mismatch by modeling semantic relevance between queries and documents, and has a great success in many retrieval tasks. We show that a deep retrieval model poorly generalizes to a new domain with large domain shift, while lexical matching and expansion models are robust across domains. To address this, we propose a simple non-parametric zero-shot hybrid model to integrate lexical matching, expansion, and deep retrieval models. Our proposed model demonstrates its effectiveness in both in-domain and out-of-domain datasets.

A recent work~\cite{Sciavolino2021} found that deep retrieval models underperform lexical models for rare entities in an entity-centric QA task. As a future work, we plan to investigate the effectivenss of our hybrid model in this task. Additionally, we plan to parameterize the hybrid retrieval model using query structure, query length, the degree of domain shift, and other signals that may reflect the performance of each individual model. Finally, we plan to explore techniques that improve the utility of out-of-domain deep retrieval models via domain adaptation.

\bibliographystyle{plain}
\bibliography{reference}

\end{document}